\documentclass[namedreferences]{SolarPhysics}
\usepackage{epsfig}

\newcommand{\be}{\begin{equation}}
\newcommand{\ee}{\end{equation}}
\newcommand{\beq}{\begin{eqnarray}}
\newcommand{\eeq}{\end{eqnarray}}

\newcommand{\aspj}[3]{  #1, {\it Astrophys. J.} {\bf  #2}, #3.}

\newcommand{\sph}[3]{   #1, {\it Solar Phys.} {\bf  #2}, #3.}

\begin{document}
\begin{article}
\begin{opening}

\title{From GHz to mHz: A Multiwavelength Study of the Acoustically  Active
14 August 2004 M7.4 Solar Flare}

\author{J.C. \surname{Mart\'inez-Oliveros}$^{1}$,
        H. \surname{Moradi}$^{1}$,
        D. \surname{Besliu-Ionescu}$^{1,2}$,
        A.-C \surname{Donea}$^{1}$,
        P.S. \surname{Cally}$^{1}$,
        C. \surname{Lindsey}$^{3}$}

\runningauthor{Mart\'inez-Oliveros,Moradi, Besliu-Ionescu,Donea, Cally, Lindsey}
\runningtitle{From GHz to mHz: Study of the 14 August 2004 M7.4 Solar Flare}

\institute{$^{1}$ Centre for Stellar and Planetary Astrophysics, School of
Mathematical Sciences, Monash University, Australia
                  \email{Juan.Oliveros@sci.monash.edu.au}\\
                $^{2}$ Astronomical Institute of the Romanian Academy, Bucharest,
Romania\\
                $^{3}$ NorthWest Research Associates, CORA Div. 3380 Mitchell Ln,
Boulder, CO 80301 USA
             }

\date{Received ; accepted }

\begin{abstract}
We carried out an electromagnetic acoustic analysis of the solar flare of 14 August 2004 in active region AR10656 from the radio to the hard X-ray spectrum. The flare was a GOES soft X-ray class M7.4 and produced a detectable sun quake, confirming earlier inferences that relatively low-energy flares may be able to generate sun quakes. We introduce the hypothesis that the seismicity of the active region is closely related to the heights of coronal magnetic loops that conduct high-energy particles from the flare. In the case of relatively short magnetic loops, chromospheric evaporation populates the loop interior with ionized gas relatively rapidly, expediting the scattering of remaining trapped high-energy electrons into the magnetic loss cone and their rapid precipitation into the chromosphere. This increases both the intensity and suddenness of the chromospheric heating, satisfying the basic conditions for an acoustic emission that penetrates into the solar interior.
\end{abstract}

\end{opening}

\section{Introduction}

\inlinecite{wolff72} suggested for the first time that solar flares would release acoustic noise into the solar interior. He even suggested that these, and perhaps comets, were the primary source of solar oscillations discovered by \inlinecite{leighton62}. We know that most of the solar oscillations visible on the Sun's surface are too short-lived to be driven by flares, even at solar maximum, and are now believed to be driven by convection \cite{stein67,stein74}. \inlinecite{kz1998} suggested that large flares might produce sun quakes to be detected against the background of solar oscillations.

 \inlinecite{kz1998} made the first identification of a
sun quake, emanating from the X2.6 flare of 9 July 1996.  The sun quake
was assumed to be the signature of intense beams of high energy
particles impinging into the lower solar atmosphere from the overlying
corona. \inlinecite{dbl1999} applied computational seismic holography to
helioseismic observations of the 9 July 1996 flare to make
high-quality seismic images of its seismic source.
Follow-up efforts to detect seismic emission from several other flares 
in 1998 and 1999, some considerably larger than the X2.6 flare of 9 July 1996, showed no other instances of significant acoustic emission \cite{Donea2004}. 
This made it evident that some flares are far more efficient emitters of 
seismic energy into the solar interior than others.

Recent developments in the study of flare acoustic emission 
\cite{kz1998,dbl1999,dl2005} encourage the view that the seismic emission 
from flares is a major discovery with a broad range of diagnostic and control 
applications for helioseismologists and flare analysts.  In order to study the effect of flares on solar oscillation modes, a
ring-diagram technique has been used by \inlinecite{ametal2003} and \inlinecite{ametal2004}. They reported an  increased power in the p-modes associated with
the flaring region. At present, it is difficult to give a
detailed comparison of these studies with the helioseismic holography 
 results and a deeper analysis is required. 

Helioseismology of ``sun quakes,'' circular waves propagating outward along 
the solar surface from an impulsive flare $\approx 30-60$~minutes after the 
impulsive phase, offers us the opportunity to explore the acoustics of 
flares themselves as well as the subphotospheres of the active regions
that produce them.

\inlinecite{dl2005} produced seismic images of the seismic sources of 
the two large flares of 28\,--\,29 October 2003.  
They suggested that photospheric heating could account for much of the seismic 
emission seen and that this may be the result of high-energy protons, which 
were evident from characteristic $\gamma$-ray signatures seen by RHESSI in the 
flares of 28\,--\,29 October 2003. 
However, the flare of 15 January 2005 (an X1.2 class flare)
showed no signature of protonic $\gamma$-rays. This  led \inlinecite{donea2006a} and \inlinecite{moradi2006} to suggest that the photospheric heating that they supposed would drive the seismic transient might
have been the result of back-warming by the downward emitted component
of intense Balmer and Paschen continuum radiated from the overlying
heated chromosphere \cite{hudson1972,metcalf2003}.

Sun quakes are not extremely rare, and they emanate from compact sources that represent only a small fraction of the energy emitted from flares. \inlinecite{dl2005} considered the possibility that relatively weak flares might be able to produce detectable sun quakes and that acoustically active flares might indeed be much more common than previously thought. This has turned out to be the case, as a comprehensive survey of helioseismic observations of flares from the Michelson-Doppler Imager (MDI) aboard the  Solar Heliospheric Observatory (SOHO) covering a significant fraction of solar activity cycle 23 by \inlinecite{donea05s} and \inlinecite{betal2006} has shown. 

Indeed, on 9 September 2001 at 20:40 UT, an M9.5 flare occurred in
active region AR\,9608. \inlinecite{donea2006a} have extensively analysed
the seismic transient of this flare.  The helioseismic signatures of
this flare drew our attention to several important points: the
acoustic signature of the flare was quite compact and was spatially
and temporally consistent with the white-light signature, reinforcing 
the suggestion that sudden heating of the photosphere may contribute 
significantly to the seismic emission detected. 
They also found that the acoustic signature was spatially and temporally 
coincident with suddenly changing magnetic signatures, suggesting that 
suddenly changing magnetic forces might have contributed to the seismic 
emission.\footnote{\inlinecite{donea2006a} and \inlinecite{moradi2006}  
have expressed concern as to whether the magnetic signatures are the result 
of real changes in the photospheric magnetic field. \inlinecite{kz2001} also reported similar magnetic signatures in flares. They expressed concerns about possible effects of an inversion of the Ni {\sc i}~6768~\AA\ line as a result of heating of the solar atmosphere by high-energy particles. \inlinecite{sh2005} likewise found transient magnetic signatures in flaring photospheres. \inlinecite{qg2003} attribute the sign reversal in the MDI magnetic signature of an impulsive flare to radiative-transfer effect. Clearly, these are concerns that need to be considered.}
The fraction of energy emitted into the subphotosphere as seismic waves 
remained a small fraction of the total energy released in the flare. 
The persistence of a sudden, co-spatial white-light signature in flares 
where no energetic protons were evident was consistent with acoustic 
emission driven by back-warming of the low photosphere by radiation 
from a heated overlying chromosphere.

In this paper, we report the discovery of a seismic transient produced by the 
M7.4 solar flare of 14 August 2004 in AR 10656. We have derived phase-coherent seismic images of the source of this flare from Doppler seismic observations of the flare by the MDI using computational seismic holography. Other supporting hard X-ray observation data included in this study are from the Reuven Ramaty High Energy Solar Spectroscopic Imager (RHESSI), soft X-ray emission from the Geostationary Operational  and Environmental Satellites (GOES), visible continuum emission from the Global 
Oscillations Network Group (GONG+), H$\alpha$ emission from the Big Bear Solar 
Observatory (BBSO) and radio emission from the Nobeyama Radio Heliograph (NoRH). 
We will compare these observations with the holographic images.

\section{The Helioseismic Observations}

The MDI data we utilised consist of full-disk Doppler images in the photospheric 
line Ni {\sc i}~6768~\AA, obtained at a cadence of 1 minute, in addition to approximately
hourly continuum intensity images and line-of-sight magnetograms.
The MDI data sets are described in more detail by \inlinecite{setal1995}.
For the flare of 14 August 2004, we analysed a dataset with a period of
4 hours around the time of the flare. We also obtained visible continuum maps of 
AR10656 during the flare from the GONG observatory at Mauna Loa. Technically, the 
GONG ``continuum intensity maps'' represent a measure of radiation in a 
$\approx$0.7~\AA\ bandpass centred on the Ni {\sc i}~6768~\AA\ line, whose equivalent 
width is 0.07~\AA .

For the purpose of our analysis, all MDI and GONG images were remapped as a Postel 
projection \cite{deforest2004} that tracks solar rotation, with the region of interest fixed at the 
center of the projection. The nominal pixel separation of the projection was 
0.002~solar~radii (1.4~Mm) with a $256\times256$ pixel field of view.

\section{The Acoustic Signatures}

AR10656 first appeared on the solar surface on 7 August 2004 at S12E55 
($-758''$,$-253''$) as an $\alpha$ sunspot. Over the next seven days, the active 
region continued to increase in magnetic complexity and evolved to a 
$\beta\gamma\delta$ type. During the period 8\,--\,16 August it produced 2 
X-class, 36 M-class and more than 150 C-class solar flares.

On 14 August, the active region was situated at S13W36 ($542''$,$-298''$)
and was characterised by a strong $\delta$ configuration in the center
of the sunspot, and an overall configuration of
$\beta\gamma\delta$. At 05:36 UT an M7.4-class solar flare occurred,
peaking at 05:44 UT and concluding at 05:52 UT (as given by GOES12)
with an X-ray flux of $3.8\times10^{-2} \mathrm{ J~m^2}$. 
This flare produced significant seismic emission, and is the least energetic 
flare in soft X-rays known to have generated a detectable acoustic transient. 

It should be emphasised that the same active region produced two other significant
seismic transients within a period of 48 hours: the first was generated by an X1.0 
flare on 13 August 2004; the second was generated by the M9.4 solar flare on 15 
August 2004 \cite{betal2006,donea2006a}.
We applied computational seismic holography to the helioseismic observations to 
image the acoustic sources of these sun quakes. 
This method is described in depth by \inlinecite{lb2000}, and has been used 
extensively in flare seismology, with great success in identifying numerous 
seismic sources from solar flares \cite{dbl1999,dl2005,donea2006a,moradi2006}.
Helioseismic holography is essentially the phase-coherent reconstruction of
acoustic waves observed at the solar surface into the solar interior to render
stigmatic images of subsurface sources that have given rise to the surface
disturbance. Because the solar interior refracts down-going waves back to the 
surface, helioseismic holography can likewise use observations in one surface 
region, the pupil, to image another, the focus, a considerable distance from 
the pupil. 
This is referred to as ``seismic holography from the subjacent vantage'' 
\cite{lb2000}. 
The subjacent vantage renders the photosphere as viewed by an acoustic observer 
directly beneath it.
In general the acoustic reconstruction can be done either forward or backward 
in time. When it is backward in time, we call the extrapolated field the 
``acoustic egression.'' In the case of subjacent vantage holography, this 
represents waves emanating from the surface focus downward into the solar 
interior.
\begin{figure*}[!ht]
\begin{center}
\includegraphics[width=1.0\textwidth,height=0.70\textheight]{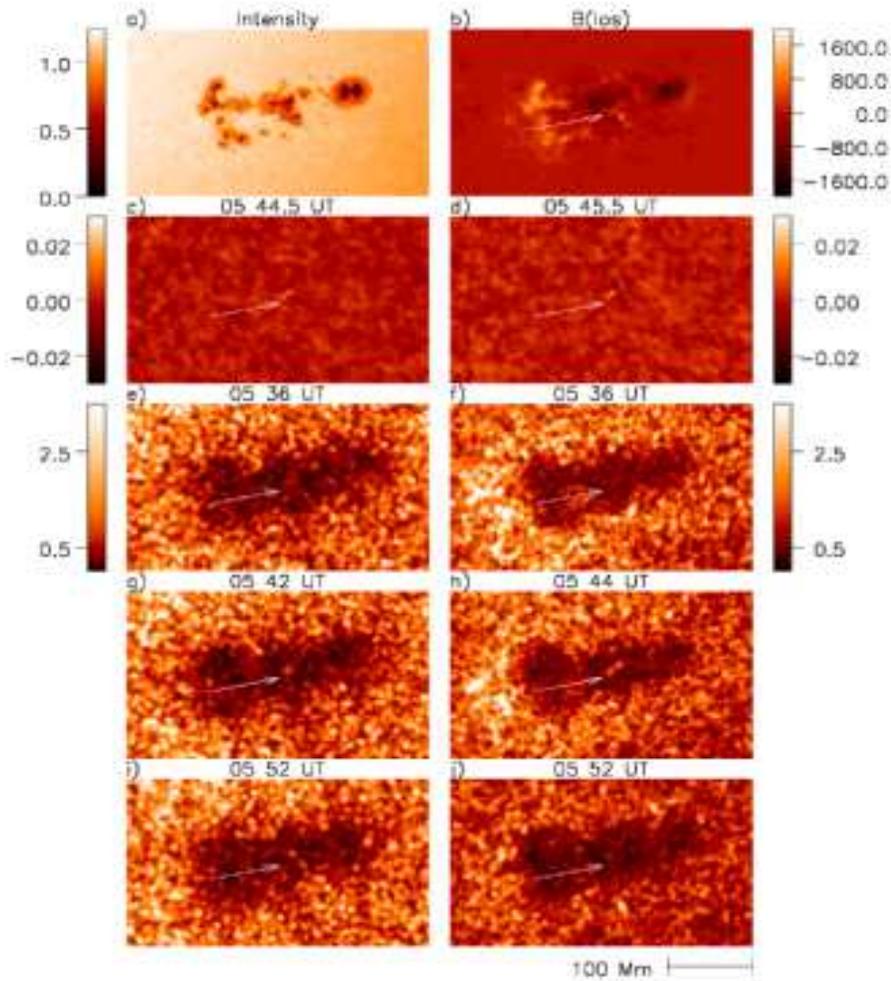}
\caption{Egression power snapshots of AR10656 on 14 August 2004 taken before,
during and after the flare and integrated over a 2.0\,--\,4.0~mHz and 5.0\,--\,7.0~mHz frequency band.
Top frames show a MDI visible continuum image of AR10656 (left) at 06:24~UT and a 
magnetogram (right) at 05:44~UT. 
Second row shows GONG continuum intensity differences 30 seconds before and after 
the time that appears above the respective frames.
Bottom three rows show egression power maps before (row 3), during (row 4),
and after (bottom row) the flare at 3~mHz (left column) and 6~mHz (right
column). 
Times are indicated above respective panels, with arrows inserted to indicate the 
location of the acoustic source. Color scales at right and left of row 3 apply to 
respective columns in rows 3\,--\,5. The seismic region is easily seen in a movie of the egression power maps. For a better visualisation of the acoustic source, we have enhanced the area of the seismic signature by a factor of 1.5.}
\label{egpwrsnaps}
\end{center}
\end{figure*}

\begin{figure*}[!ht]
\begin{center}
\includegraphics[width=1.0\textwidth, angle=0]{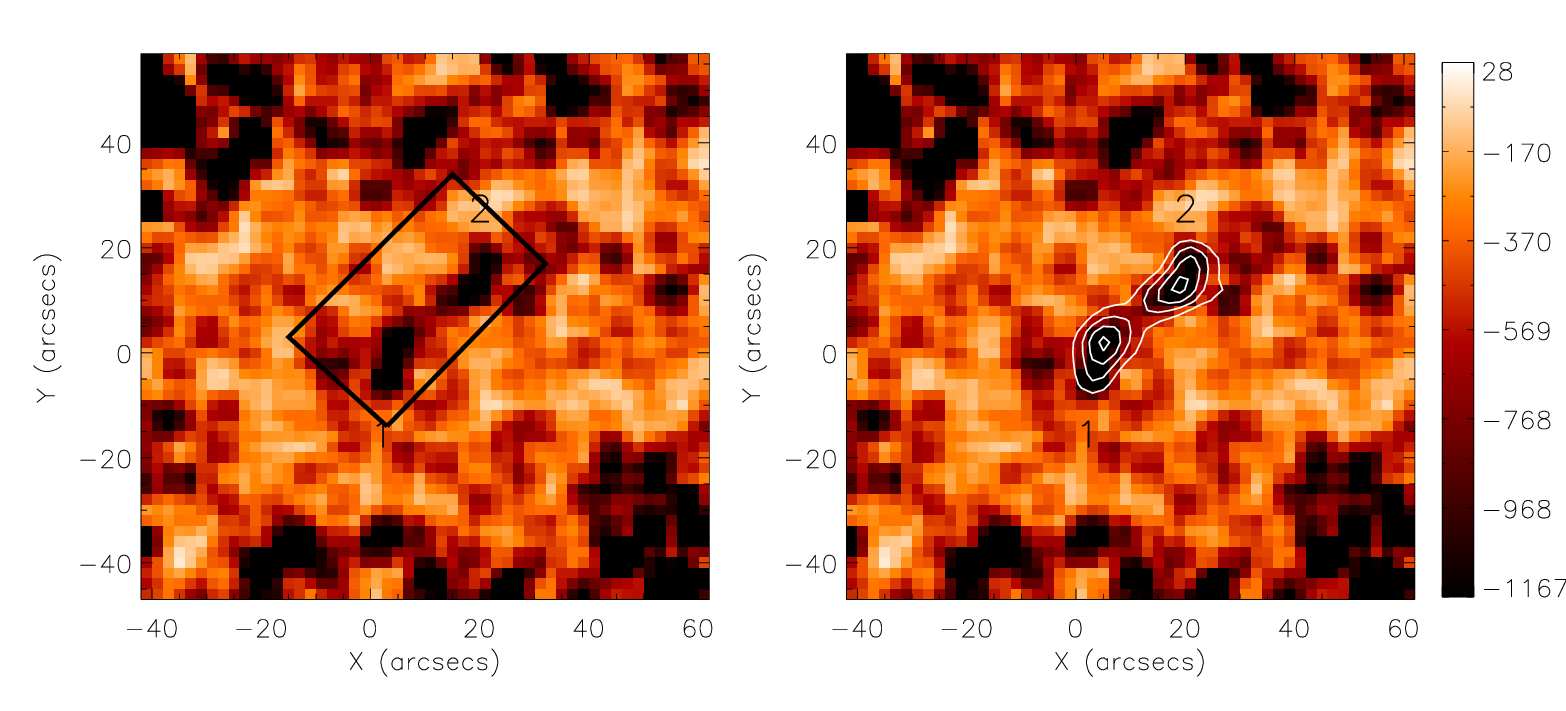}
\caption{A zoom image of the 6 mHz egression power snapshot seen in Figure \ref{egpwrsnaps}(h) taken at 05:44 UT. The color map of the image was inverted for a better visualisation of the acoustic source morphology. The left panel shows the acoustic kernels (labelled 1 and 2) and the right panel 
shows the same image but with egression power contours overlaid. The acoustic 
source 1 appears to be the stronger of the two. The rectangle represents the seismic region which we used in this paper to study the time series.}
\label{egpwrmag}
\end{center}
\end{figure*}

To assess seismic emission from the flare, we computed the egression over the 
neighborhood of the active region at 1-minute intervals, mapping the egression 
power for each minute of observation. The resulting egression power movies and 
``snapshots'' (egression power sampled over the solar surface at any definite 
time) are computed over 2~mHz bands, centred at 3~mHz and 6~mHz. 
The higher frequency band has a number of advantages in that it avoids the much 
greater quiet-Sun ambient noise that predominates the 2\,--\,4~mHz frequency 
band and due to a shorter wavelength, it also provides us with images that have 
a finer diffraction limit. 
However, these advantages come at some expense in temporal discrimination, as 
the temporal resolution of egression computations is limited to
\begin{equation}
\Delta t ~=~ \frac{1}{\Delta\nu} ~=~ \frac{1}{2~{\rm mHz}} ~=~ 500~{\rm s}.
\label{eq:Deltat}
\end{equation}
This temporal smearing results in the acoustic signature of the flare commencing 
several minutes before the actual onset of the flare and lasting for several 
minutes afterward, even if the actual acoustic disturbance was instantaneous.

\begin{figure*}[!ht]
\begin{center}
\includegraphics[width=1.0\textwidth]{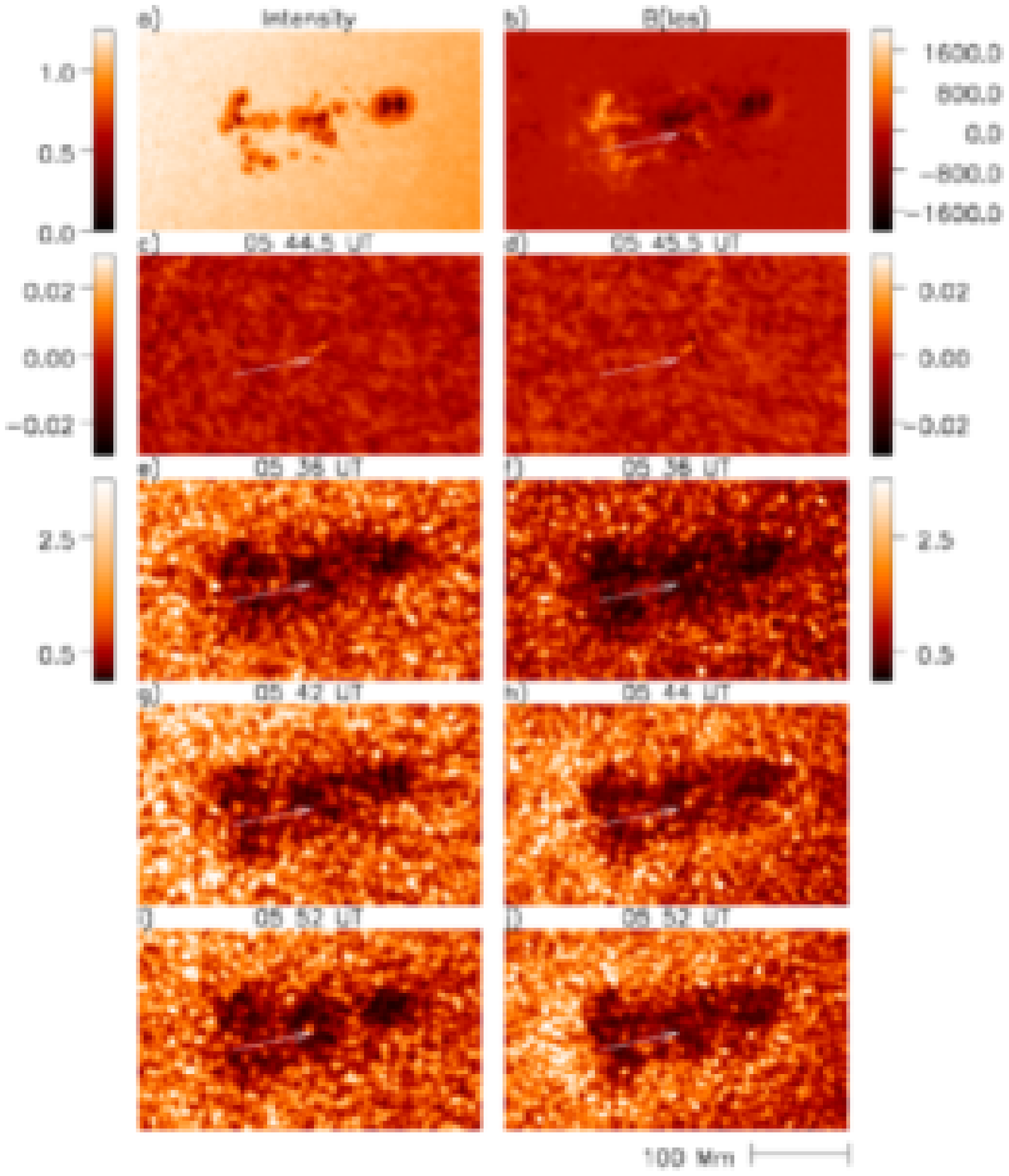}
\caption{Acoustic power snapshots of AR10656 on 14 August 2004. 
Details are the same as for Figure~\ref{egpwrsnaps}, but local acoustic power maps 
appear in the bottom three rows in place of egression power maps.}
\label{acpwr_snaps}
\end{center}
\end{figure*}

Egression power snapshots before, during and after the flare are shown in
the last three rows of Figure~\ref{egpwrsnaps} at 3~mHz (left column)
and 6~mHz (right column).
In these computations the pupil was an annulus of radial range 15\,--\,45~Mm
centered on the focus. 
To improve the statistics, the original egression power snapshots are smeared by 
convolution with a Gaussian with a $1/e$-half-width of 3~Mm. 
The egression power images and the continuum images are also normalised to unity 
at respective mean quiet-Sun values.
At 3~mHz this is $\sim$2.0~kW~m$^{-2}$.
At 6~mHz it is 70~W~m$^{-2}$.
\begin{figure*}[!ht]
\begin{center}
\includegraphics[width=0.82\textwidth, angle=0]{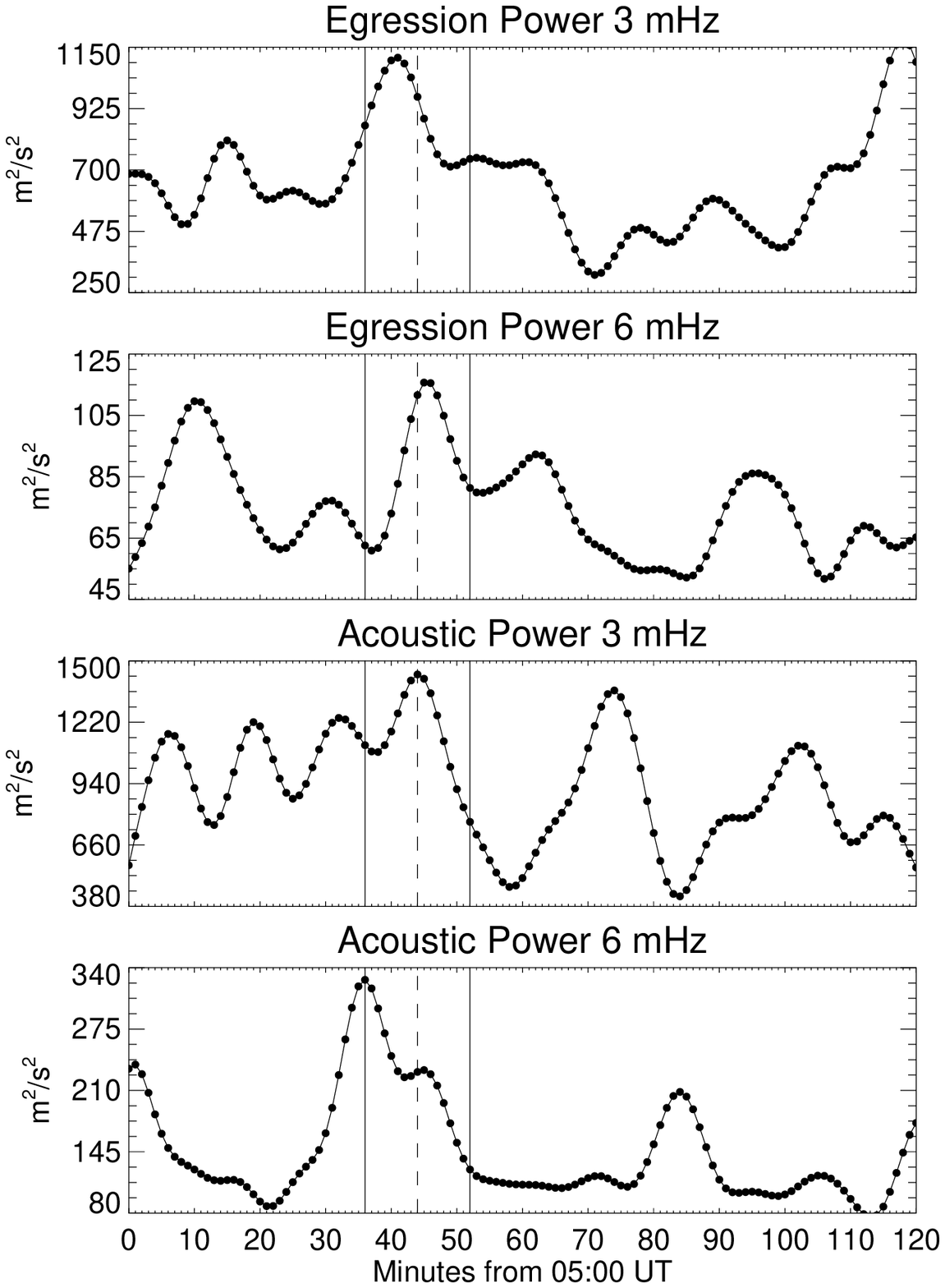}
\caption{3 and 6 mHz egression power and acoustic power time series, integrated over 
the neighborhood of the egression power signatures, are plotted in the top four 
rows. 
The vertical lines represent the beginning (05:36 UT), maximum (05:44 UT) and ending 
(05:52 UT) times of the GOES X-ray flare. The relatively extended duration of the 
acoustic signatures is a result of limits to temporal resolution imposed by 
truncation 
of the spectrum (see equation \ref{eq:Deltat}).}
\label{timeseries}
\end{center}
\end{figure*}

All egression power snapshots mapped in Figure~\ref{egpwrsnaps} show
considerably suppressed acoustic emission from the magnetic region, attributed 
to strong acoustic absorption by magnetic photospheres, discovered by 
\inlinecite{bdl1988} (see also \opencite{braun1995}; \opencite{blff1998};
\opencite{bl1999a}). 
Furthermore, all 6~mHz egression power snapshots in Figure~\ref{egpwrsnaps}  
show acoustic emission ``halos,'' {\it i.e.} significantly {\it enhanced} acoustic 
emission from the outskirts of complex active regions \cite{bl1999b,dbl1999}.

Looking at Figure~\ref{egpwrsnaps}, a significant excess of acoustic emission 
is evident at 05:44~UT in the 6~mHz egression power snapshot, indicated by an 
arrow in all of the frames, appearing to lie across the penumbral magnetic 
neutral line and spanning $\approx$25~Mm in length. Upon closer inspection, we 
can see from the zoomed egression power snapshot in Figure~\ref{egpwrmag}, that 
there are in fact two separate components to the seismic source (acoustic 
kernels) that appear to be separated by $\approx$7~Mm when they initially appear 
(05:39~UT), and because of their close proximity and evolution with time, they 
seem to appear as one extended source in Figure~\ref{egpwrsnaps}. 
These acoustic kernels coincide closely with hard X-ray (HXR) signatures (see 
Section \ref{hxrradio} and Figure~\ref{smeared}), indicating that high-energy 
particles accelerated above the chromosphere contribute to the excitation of 
the seismic source. The egression power map in Figure~\ref{egpwrmag} is smeared by a factor of 0.004, in order to emphasise the source geometry and the acoustic kernels. The map also shows kernels that we associate with the fluctuating acoustic noise of the active region.

The source geometry also closely corresponds with other compact manifestations 
of the flare including significant white-light emission with a sudden onset, as 
indicated by the intensity difference signatures shown in the second row of 
Figure~\ref{egpwrsnaps}, and  microwave emission at 17 and 34 GHz.  
The 3~mHz egression power snapshots (Figure~\ref{egpwrsnaps}) also shows emission 
during the flare. 
In fact, from the egression and acoustic power time series of Figure~\ref{timeseries}, 
it appears that we have a distinct and considerably stronger seismic emission at 
3~mHz than at 6~mHz. This is because of a much greater ambient acoustic noise at 3~mHz which renders the 
considerably greater 3~mHz seismic emission signature no more conspicuous than 6~mHz.

Figure~\ref{acpwr_snaps} shows the local acoustic power snapshots of AR10656 at 
3~mHz (left column) and 6~mHz (right column) before, during and after the flare. 
Each pixel in a local acoustic power map represents the local surface motion as 
viewed directly from above the photosphere, which should not be confused with 
the egression power computed by subjacent vantage holography of the surface, 
where each pixel is a coherent representation of acoustic waves that have 
emanated downward from the focus, deep beneath the solar surface, and re-emerge 
into a pupil 15\,--\,45~Mm from the focus.
As in the case of the 6~mHz egression power, the local acoustic power maps show 
a broad acoustic deficit marking the magnetic region and an enhanced local 
acoustic power halo surrounding the active region which is also clearly apparent. 
The acoustic source is difficult to distinguish in either the 3 or 6~mHz 
acoustic power signatures.

\section{Analysis and Results}
\subsection{White Light Flare signature}

\begin{figure*}[!ht]
\centerline{\includegraphics[width=1.0\hsize]{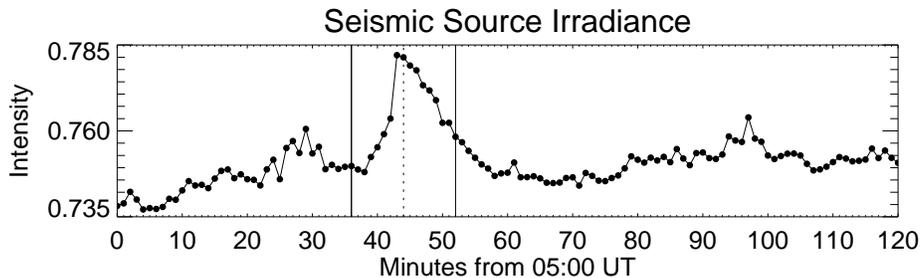}}
\caption{Time dependence of the visible continuum irradiance normalised
to the quiet-Sun and integrated over the area of the seismic source. 
The vertical lines show the flare times as in Figure~\ref{timeseries}. 
The maximum emission in white light continuum temporally coincides with the 6 mHz 
seismic emission at its maximum.
}
\label{whitelight}
\end{figure*}

Figure~\ref{whitelight} shows the  time dependence of the visible continuum 
irradiance normalised to the quiet-Sun and integrated over the area of the 
seismic source. At 05:39~UT the irradiance began to increase for $\sim$4~minutes, 
then underwent a sudden jump at 05:42 UT for approximately 2 minutes 
and then slowly decreased to the background level. The maximum irradiance was 
approximately 4\% above the quiet-Sun mean.

The white light flare signature is spatially co-aligned with the emission of the 
seismic sources as imaged in Figure~\ref{egpwrsnaps}.

\subsection{The Magnetic Field}

\inlinecite{schunker2005} have shown that magnetic forces are of particular significance 
for acoustic signatures in penumbral regions, where the magnetic field is
significantly inclined from vertical. 
Therefore, understanding the 3-D magnetic configuration of the coronal loops 
hosting flares would give us a powerful control utility for seismic diagnostics 
of active region sub-photospheres.  
This will be useful for addressing questions concerning the MHD of inclined magnetic 
fields, the role of fast and slow magneto-acoustic mode coupling in magnetic 
photospheres, sub-photospheric thermal structure, and how wave generation by 
turbulence in active region sub-photospheres differs from that in the quiet
sub-photosphere.

\begin{figure*}[!ht]
\centerline{\includegraphics[width=1.0\hsize]{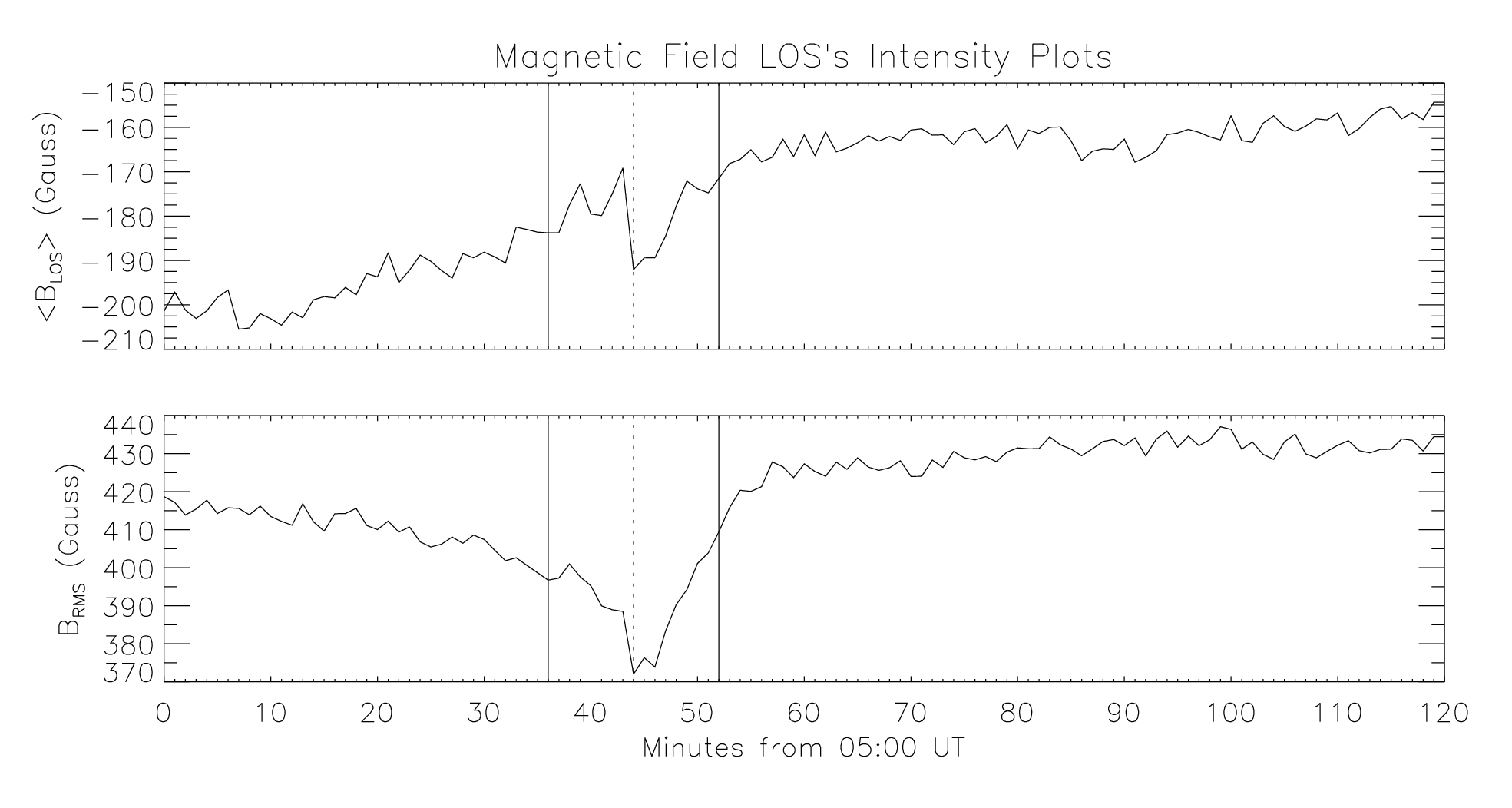}}
\caption{Time series of the mean and the root mean square of the LOS magnetic 
field integrated over the area of the seismic source.
}
\label{mag}
\end{figure*}

In Figure~\ref{mag} we have shown the time series of the mean and the root mean 
square (RMS) values of the line-of-sight (LOS)  magnetic field, 
integrated over area of the seismic source (the integration area is plotted in Figure~\ref{egpwrmag} - black rectangle - and its area has a value of $\approx$247 $\mathrm{Mm^2}$). 
The vertical lines mark the time frame of the flare.

The mean LOS magnetic field shows a steady increase from 05:10 to
06:00 UT with a strong variation as a sudden decrease, at the maximum
of the flare (05:44 UT).  The RMS of the magnetic field intensity
shows a sudden decrease of about 9.6\% of the background level, and a
sudden recovery to a 3.6\% increased background, as compared to the
background level before the flare (similar changes have been observed
by \opencite{kz2001}; \opencite{sh2005}; \opencite{ametal1993};
\opencite{wangetal2005}; see also the footnote~1 for more
references.). To obtain a general idea of the configuration of the coronal
magnetic field lines in AR10656 we computed the potential field
extrapolation by applying the code described in \inlinecite{sakurai}
to the MDI line-of-sight magnetogram. According to this extrapolation
(Figure~\ref{potential}, top frame), the field lines whose footpoints were
planted in the general region of the acoustic emission were relatively
low and compact, suggesting that the magnetic loops, into which 
particle acceleration occured during the reconnection,  were
relatively short. The second panel in Figure 9 shows the appearance after the flare maximum of more magnetic field
lines connecting the positive and negative polarities. A small
difference in the line-of-sight magnetic field configuration in the
region of the acoustic emission described by the inclined rectangle
is also noticeable.

\subsection{The Surface Ripple}

\begin{figure*}[!ht]
\centering \includegraphics[width=1.0\textwidth]{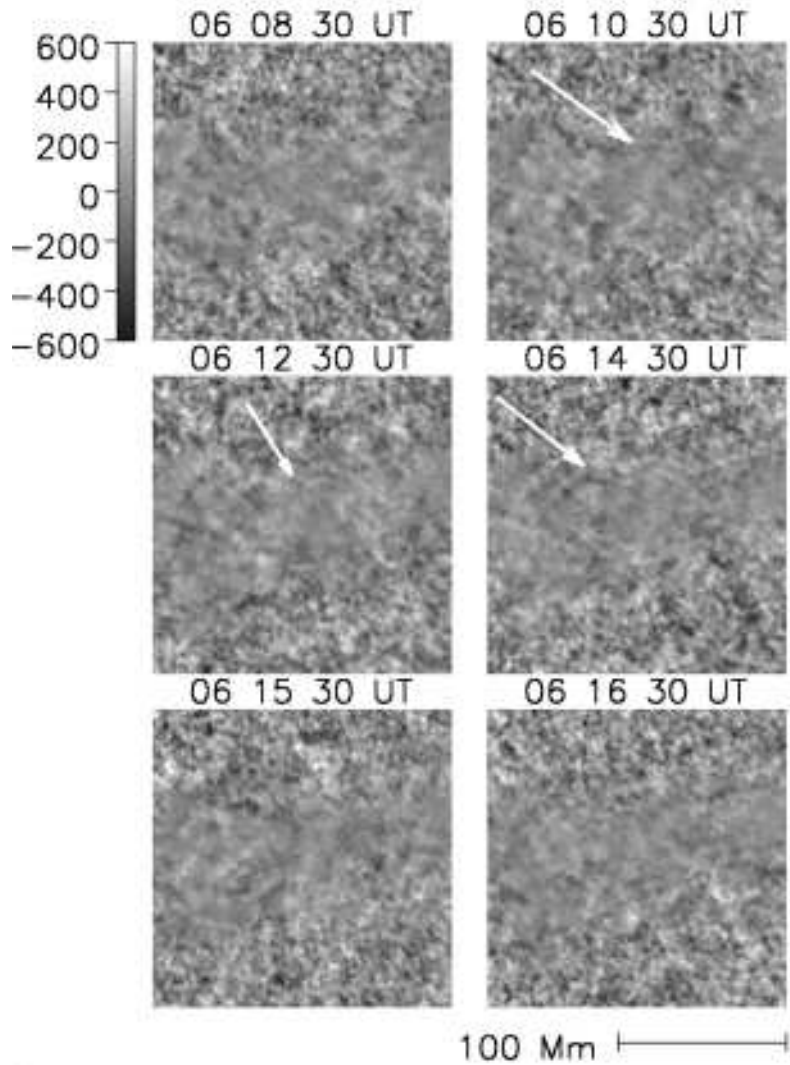}
\caption{Observations of surface ripples at the specified times
emanating from AR10656 following the impulsive phase of the
flare. Arrows show the location of the surface ripples. Only the north
angular sector of the ripple can be seen by eye.}
\label{ripple_fig}
\end{figure*}

We computed differences between consecutive Doppler frames, separated
by one minute in time, around the time of the flare to reconstruct
time-distance profiles of this seismic emission.  In this sequence we
see a surface ripple propagating in the North direction, over the
range $-$50$^\circ$ to $+$20$^\circ$ from due north in a reference
frame centered on the seismic source.  The surface ripple represents
acoustic waves that propagated tens of Mm into the solar interior from
the acoustic source and were refracted back to the surface
30~minutes after the impulsive phase of the flare. Because of
the strong fluctuating motions of the background, the ripple is
difficult to see in individual dopplergrams. They are easily recognised
in a movie of differences of consecutive Doppler frames. Even so, we
are able to see the ripple at approximately 06:10\,--\,06:15~UT. The
arrows in Figure~\ref{ripple_fig} indicate their location.  The
ripples expand into the north quiet Sun before becoming submerged into the
ambient noise.
%
%
We do not see an expanding wave moving southward, either because the
signal is too weak to be detected by eye or the emission to the north
is simply stronger. The seismic wave is highly anisotropic, its
amplitude varies with angle. The strongest amplitude is observed in
the north direction. In section \ref{hxrradio} we will see that this
direction is also approximately the direction of the motion of HXR
footpoints. A similar behaviour was reported by \inlinecite{dl2005} in the  seismically active flares of the October 2003. The fronts
of the eastern, southern and western acoustic seismic wave propagate
through the sunspot, and are exposed to a strong local magnetized
environment. As a result a significant decay and some distortion is expected, weakening the surface ripple.

\begin{figure*}[!ht]
\centering
\includegraphics[width=0.95\textwidth,height=7.5cm]{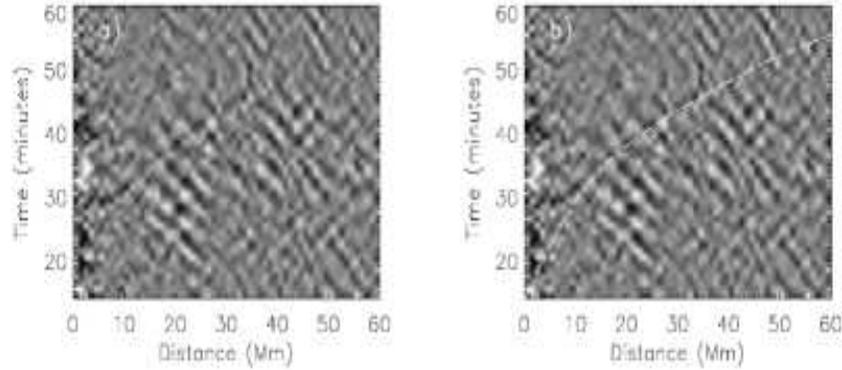}
\caption{Time-distance propagation amplitude of the surface ripple averaged over
curves of constant radius, over azimuths from $-$50 to $+$20$^\circ$ is 
rendered in gray tone in both frames. 
The curve superimposed in the right frame represents the wave travel time, $t$, for 
a standard model of the solar interior.
The time 05:30 UT is represented by $t = 0$.}
\label{td_fig}
\end{figure*}

\begin{figure*}[!ht]
\centering
\begin{tabular}{c}
\includegraphics[width=0.60\textwidth]{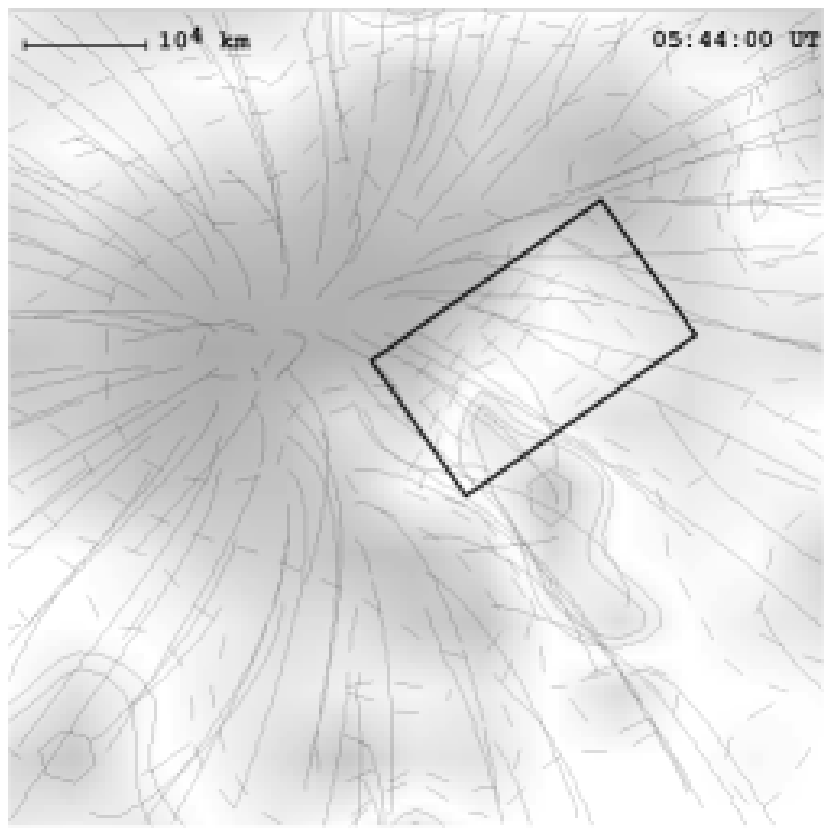} \\
        \\
\includegraphics[width=0.60\textwidth]{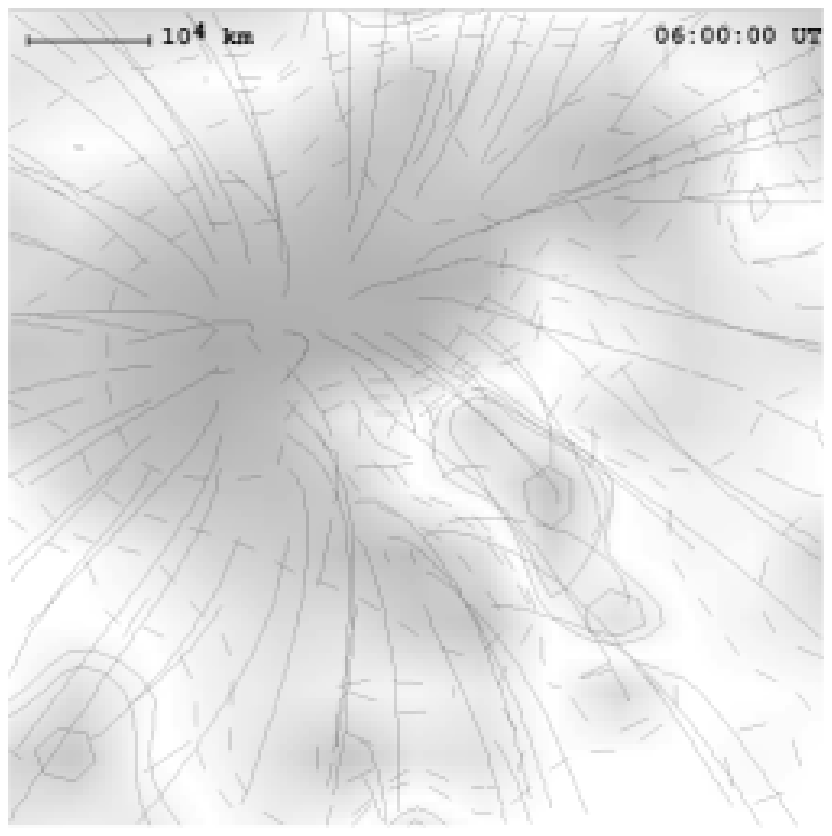}
\end{tabular} 
\caption{Potential magnetic field extrapolation of SOHO-MDI magnetograms. Top: Magnetic field extrapolation at 05:44:00~UT. Bottom: Magnetic field extrapolation at 06:00:00~UT. The grayscale background image shows the absolute value of the line-of-sight magnetic field. The dashed lines represent the negative magnetic polarity, while the solid lines represent the positive magnetic polarity. The contour lines levels are 50, 100, 300, 500, 1000 G. In the image North is up, the dimension are 104 by 104 arcsec centred at ($462$,$-303$) arcsec.}
\label{potential}
\end{figure*}

Figure~\ref{td_fig}a shows a time-distance amplitude profile for the ripple described above. The Doppler difference amplitude was averaged along curves of constant radius in the reference frame described above over the $-$50 to $+$20$^\circ$ range of azimuths over which the surface ripple was visible.
This resulting gray-tone plot is shown in Figure~\ref{td_fig}(b) with the theoretical group travel time plotted for reference.%
\footnote{This travel time, $t(\rho)$, is defined by the path integral
\begin{equation}
t(\rho) ~=~ \int _{\Gamma (\rho)} {ds \over c},
\label{eq:traveltime}
\end{equation}
where $\Gamma$ represents the path of least time through the quiet subphotosphere 
connecting surface points separated by an angular distance $\rho$ along the surface.}


\subsection{Radio and HXR Emission}
\label{hxrradio}
The flare of 14 August 2004 was observed with the Nobeyama Radio Heliograph 
(NoRH), at 17 GHz and 34 GHz, and the Reuven Ramaty High Energy Solar Spectroscopic 
Imager (RHESSI). Unfortunately, the totality of the impulsive and main phases of the 
flare was not observed by RHESSI, and as a result, images and time profile of the 
hard X-ray (HXR) emission just prior to, and after the maximum of the flare, are 
not available.

\begin{figure*}[!ht]
\centering
\includegraphics[width=0.8\textwidth]{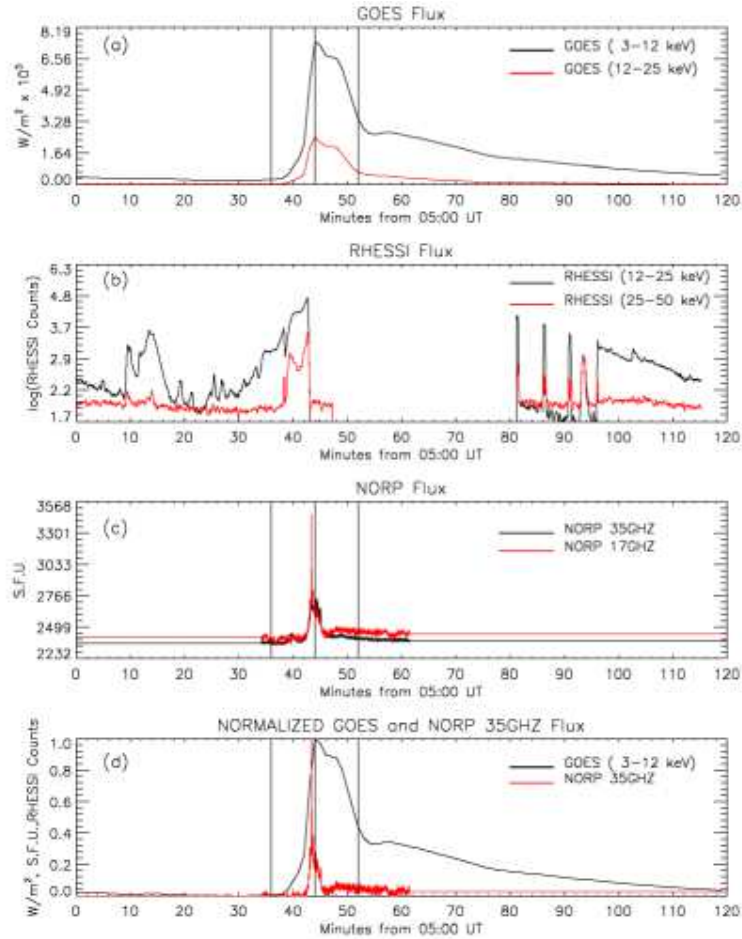}
\caption{Integrated flux time profiles: a) GOES soft X-ray 1\,--\,8~\AA\ and 
0.5\,--\,4~\AA\ channels; b) RHESSI time profiles in two channels 
12\,--\,25~keV (black line) and 25\,--\,50~keV (red line); c) Microwave time 
profiles at 17 (red line) and 35 GHz (black line); d) Normalised total GOES 
total flux (black line) and microwave flux at 35~GHz (red line). 
The vertical lines show the beginning, maximum and end of the event.}
\label{fluxes}
\end{figure*}

Figure~\ref{fluxes} shows the total flux time profiles of the event in
microwaves, soft and hard X-rays. The GOES total fluxes in the two
channels 1\,--\,8~\AA\ and 0.5\,--\,4~\AA\ are shown in the top graph
of Figure~\ref{fluxes}. Figure~\ref{fluxes}b shows the HXR-RHESSI time
profile in the two channels 15\,--\,25~keV (black line) and
25\,--\,50~keV (red line). Figure~\ref{fluxes}c shows the microwave
time profiles obtained using the Nobeyama Radio Polarimeter (NoRP)
data at 17~GHz (red line) and 35~GHz (black line). In
Figure~\ref{fluxes}d, we plotted the normalised total GOES flux at
1\,--\,8~\AA~ and the NoRP flux at 35~GHz. The empirical relation
observed between the soft X-rays flux and the HXR or microwaves is
called the Neupert effect~\cite{neupert}. It is clear from 
Figure~\ref{fluxes}d, that this effect is present and that the NoRP 35
GHz emission lags behind the GOES soft X-ray by 43 seconds. The microwave emission did not present a significant
thermal component, suggesting relatively inefficient trapping of the
accelerated electrons in the coronal magnetic field.  This result is
of significant importance to the process of transportation of energy
from the reconnection site into the lower layers of the chromosphere
and further into the photosphere where the sun quake was produced.

It has already been established that a close relationship exists between HXR 
and radio fluxes in the impulsive phase of a flare (see \opencite{kundu1}; \opencite{bastian1}). 
Based on this relationship, it is generally believed that essentially the same 
population of energetic electrons is responsible for both HXR and radio emission. 
The radio emission is thought to be produced by accelerated nonthermal electrons 
orbiting magnetic field lines and trapped in the coronal magnetic field. 
The hard X-ray emission is produced by Coulomb collisions of these energetic
electrons with the dense chromospheric plasma.

\begin{figure*}[!ht]
\centering
\includegraphics[width=0.95\textwidth]{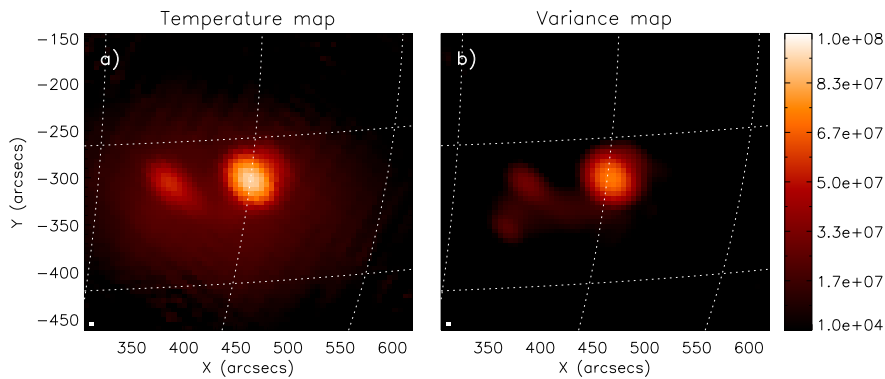}
\caption{Temperature and variance maps. 
Panel (a) shows the brightness temperature radio map at 17GHz and panel (b) shows the variance map, which identifies a non-thermal radio source at 
the location of the main spot of the AR10656.}
\label{radio}
\end{figure*}

The maximum brightness temperature of the radio source at 17~GHz (Figure~\ref{radio}, 
left panel) was measured to be 4.67$\times$10$^7$~K, with a spectral index, $\delta$,
of $-$3.67. 
These results indicate that a non-thermal emission process for the microwave 
radiation is at work; the non-thermal emission region was also confirmed using 
the variance technique%
\footnote{We calculate a variance map of a set of radio images using the following equation: 
\begin{equation}
 \sigma^2_{ij} = \frac{1}{N}\sum^N_{k=1} x^2_{ijk} - \frac{1}{N^2} \left(\sum^N_{k=1}x_{ijk}\right)^2
\end{equation} 
\noindent
where $i =1,2,\ldots,L$ is the image row number, $j=1,2,\ldots,M$ the column number and $k=1,2,\ldots,N$ is the image number in the data set.} 
for solar radio image analysis \cite{grechnev2003}. This technique allows us to plot a radio map of the non-thermal emission from the active region by also subtracting any contribution from thermal sources in the corona. From the variance map (Figure~\ref{radio}, right panel) we infer that the non-thermal emission is compact and well correlated with the HXR emission region. The flux of electrons with energies $\gtrsim$25~keV is very small, $\approx$6\% 
of the flux registered in the 12\,--\,25~keV energy band, and possibly did not make
a significant contribution to the seismic emission. A delay of 43~seconds is  observed between the microwave emission (05:43:17~UT) and the maximum in the seismic signature (05:44:00~UT).  A similar delay is observed between the NoRP 35 GHz emission and the GOES soft X-ray.

\begin{figure*}[!Ht]
\centering
\includegraphics[width=1.0\textwidth]{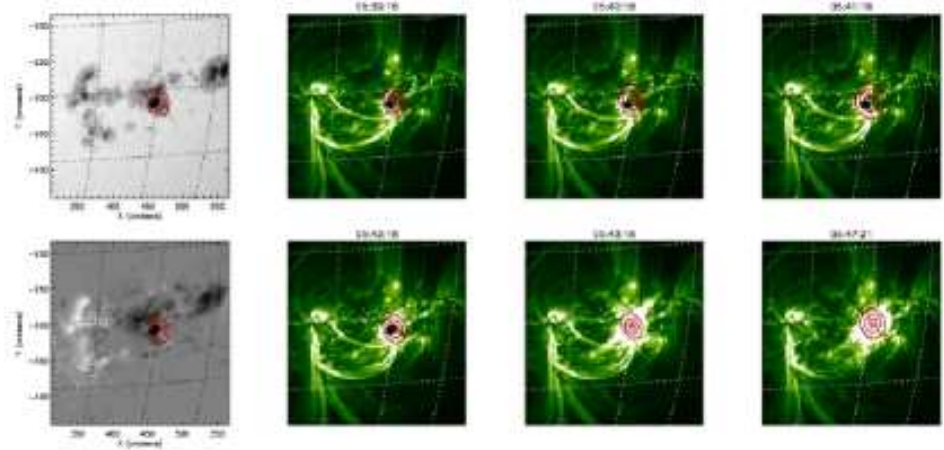}
\caption{First column shows the MDI intensity continuum and magnetogram images of 
AR10656 with the microwave emission at 17 GHz (red large contours) and RHESSI 
12\,--\,25~keV (black small contours) overplotted. Evolution of the flare at 
171{\AA} as observed by TRACE is shown in the last three columns for the specified 
times. 
RHESSI 12\,--\,25~keV HXR emission (black contours) with contour levels of 50\%, 
80\% and 95\% of the maximum source intensity, and NoRH microwave emission at 
17~GHz (red contours) at 20\%, 50\%, 80\% and 95\% of the maximum intensity of 
the radio source are also shown. 
The field of view is $256''\times256''$ with north is upward.}
\label{trace}
\end{figure*}

\begin{figure*}[!ht]
\centering
\includegraphics[scale=0.85]{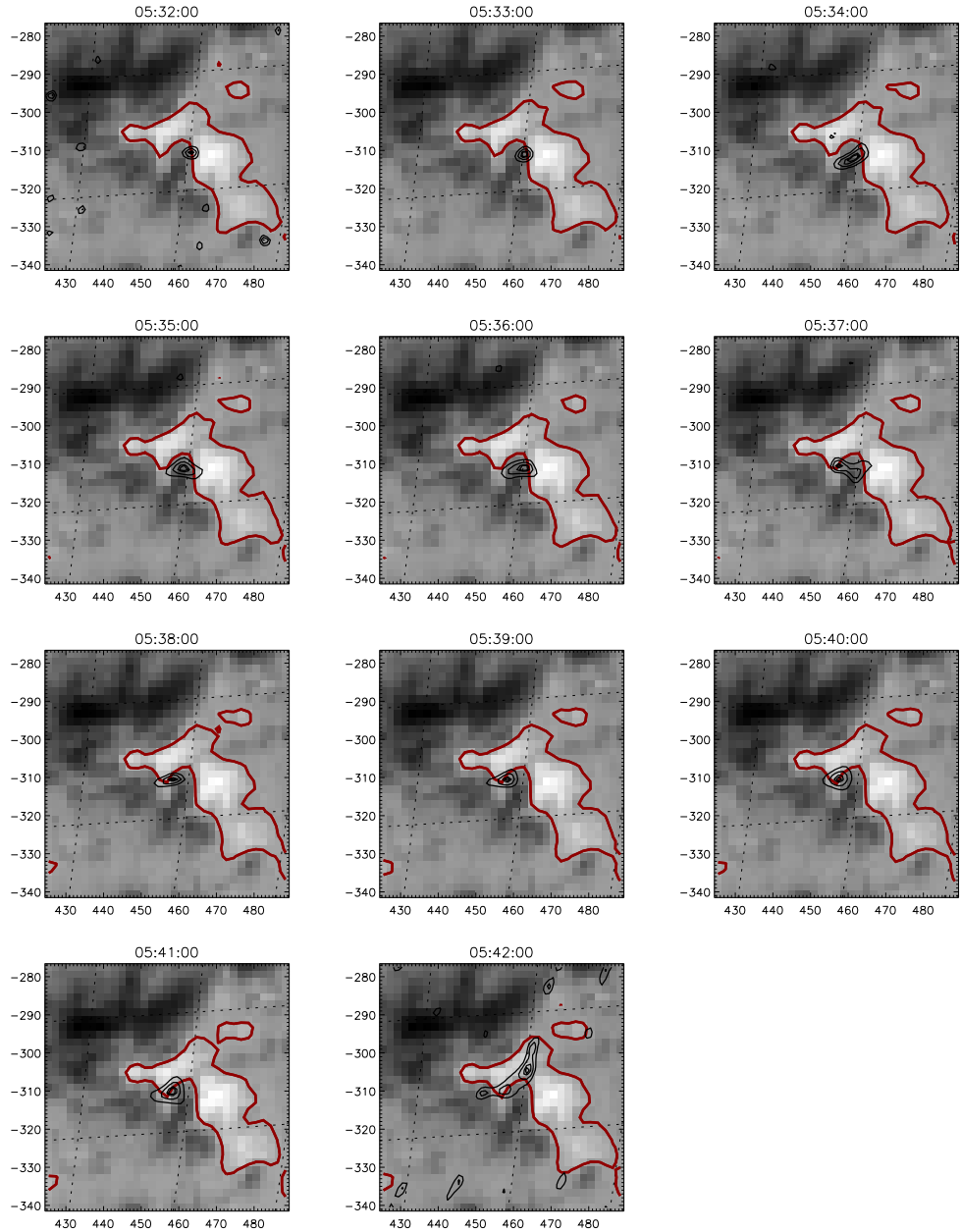}
\caption{Evolution of the MDI magnetogram (background) and RHESSI-HXR source 
12\,--\,25~keV (black contours, with levels 50\%, 80\% and 95\% of the maximum 
source intensity) from 05:32:00 UT to 05:42:00. The red line is the magnetic 
neutral line of the MDI magnetogram. North is up and East to the left.}
\label{xzoom}
\end{figure*}

Figure~\ref{trace} shows a sequence of images of the 14 August 2004 flare 
taken by the Transition Region and Coronal Explorer (TRACE) overplotted with 
the contours of the NoRH microwave emission at 17~GHz (large red contours) and 
RHESSI 12\,--\,25~keV HXR (small black contours). We applied the MEM-SATO algorithm \cite{satoal1999} available in the standard RHESSI software to reconstruct RHESSI images from grids 3,
4, 5, 6, and 8 using an integration time of 1 to 4 s. 
As seen in Figure~\ref{trace}, the impulsive phase of the flare has a simple compact 
morphology in both HXR and microwaves until the minute prior to the flare maximum, 
when the HXR source evolves into an extended source composed of three smaller kernels 
(as seen in Figure~\ref{xzoom}). 
The radio source maintained its morphology after the flare maximum. 
The close temporal and spatial correlation between the microwave and X-ray 
emissions in this flare indicates a sudden energy deposition into the chromosphere 
by non-thermal electrons. This is in agreement with the prediction made by \inlinecite{kz1998}.
Figure~\ref{trace} also shows a spatial correlation between the flare region 
observed by TRACE at 171~\AA\ and the microwave and HXR sources.

\begin{figure*}[!ht]
\centering
\includegraphics[width=0.95\textwidth]{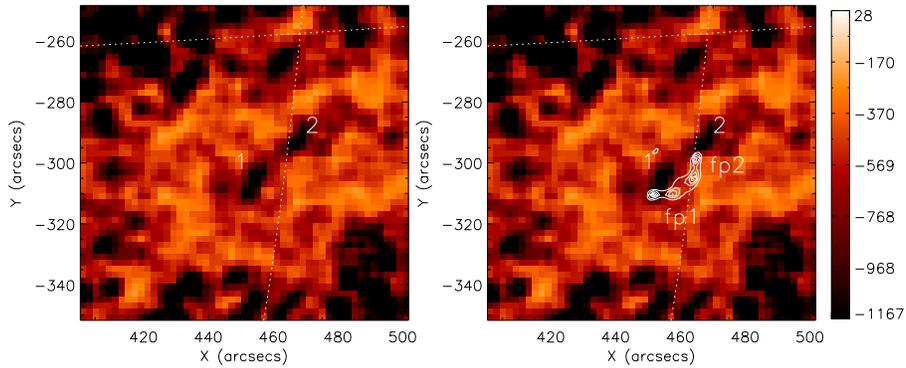}
\caption{Left: Egression power map at 6~mHz, with contour levels of 50\%, 65\%, 
80\% and 95\% of the maximum source intensity. Right: Egression power map and 
RHESSI contour plots, with levels of 30\%, 50\%, 70\% and 90\% of the maximum 
source intensity. The color map of the egression power map was inverted for a better visualisation of the acoustic source.}
\label{smeared}
\end{figure*}

The temporal evolution of the HXR feature, with respect to the photospheric magnetic neutral line, can be seen over a sequence of MDI magnetograms taken around the time of flare-maximum (Figure~\ref{xzoom}). The HXR footpoint appears to be moving in the north\,--\, north\,--\,east direction, a motion which is not parallel to the photospheric neutral line. Furthermore, we can clearly see that the source maintains its compact HXR structure until the last 
minute of observation (05:42~UT), reinforcing the observations shown in Figure~\ref{trace}. 
In this last minute, the source appears to evolve into an elongated shape that 
covers both magnetic polarities lying around the neutral line. 
This new elongated source is composed of three kernels, two of which are located 
in the positive magnetic region and the third one is located near the final 
position of the compact source observed at 05:42~UT. We remark that the motion and evolution of the RHESSI source is seen as projected over the egression power maps. The frames in Figure~\ref{xzoom} show first a loop-top emission (compact kernel) which gradually moves towards the footpoints along a single magnetic loop, the one that hosted the seismically active flare. The break-up of the HXR emission kernels began at 05:42~UT. After this time, no RHESSI data where available, but following a similar study done by \inlinecite{dl2005}, we can predict that the RHESSI footpoints and the seismic source will match in the following two minutes.
Figure~\ref{smeared} shows that the egression power snapshot at 6~mHz and the 
HXR sources have a similar morphology, with two of the four HXR sources (fp1 and fp2 in Figure~\ref{smeared}) having a strong spatial correlation with the acoustic kernel sources in the egression power snapshot.

\section{Conclusions}

The detection of seismic transients from the M-class flares opens a new era of studying seismically active solar regions. Acoustically active flares are the most compact, most impulsive, and highest-frequency solar acoustic sources discovered to date. Moreover, they are the only known sources of acoustic waves that operate in the outer, visible, solar atmosphere. This makes the transients they release into active region sub-photospheres understandable in a way that wave generation by sub-photospheric convection is not.

We carried out a study of the M-class flare of 14 August 2004 from AR10656, including HXR emission, seismic emission into the solar interior in the 2.5\,--\,7.0~mHz spectrum and radio emission up to 34~GHz. We applied holographic and other standard time-distance diagnostics to helioseismic observations of the seismic transient emitted by the flare. 
These clearly show the signature of an expanding wave packet centered on a source of HXR emission. The holographic images show a seismic source morphology composed of two kernels approximately perpendicular to the magnetic neutral line of the active region in the penumbra of one of the sunspots. The kernels are spatially aligned close to similar HXR kernels in the 12\,--\,25~keV energy range. Visible continuum emission, similarly aligned with the holographic kernels, reinforces the hypothesis, based on similar instances in other seismically active flares, that heating of the photosphere contributes to the observed seismic emission, possibly as a result of back-warming by the chromospheric source of the continuum emission.

The loss of HXR observations before HXR maximum encumbers our ability to conduct 
a realistic comparative analysis based on timing. Nevertheless, a simultaneous rise 
in the HXR flux with the 17~GHz and 34~GHz radio flux  suggests that roughly the 
same particles, relativistic electrons, produce both the radio and HXR emission. 
The radio signature, attributed to gyro-synchrotron emission from relativistic 
electrons, is highly impulsive, both at the onset and the ensuing decline phases.

Gyro-synchrotron emission from flares is often characterized by an impulsive rise
followed by a rapid but sometimes only partial decline in brightness temperature.
Then follows a slow decline of the remaining signature over many minutes.
The latter behavior is broadly attributed to electrons that are trapped 
in a magnetic flux tube because they were injected into the tube in a direction 
that lies outside of the magnetic loss cone \cite{kundu1}. 
These electrons may be scattered into the loss cone by ambient thermal electrons 
in the flux tube and leak into the chromosphere over a duration that depends
on the scattering rate, which in turn depends on the density of ambient thermal
electrons in the flux tube.
Whether these temporarily trapped electrons can contribute to seismic emission 
depends on the foregoing duration, since a significant contribution to the seismic 
transient is thought to depend critically on thick target heating that is relatively 
sudden, within about a minute or so.
A rapid increase in the thermal free electron and ion density due to ablation 
of the upper chromosphere might facilitate the rapid injection of initially 
trapped relativistic electrons into the loss cone significantly increasing both 
the magnitude and suddenness of chromospheric and photospheric heating thought 
to contribute to seismic emission.
Chromospheric ablation into the magnetic flux tube by relativistic electrons 
initially injected into the loss cone can greatly enhance scattering by ambient 
electrons and ions in the magnetic flux tube, if the flux tube is filled with
this material sufficiently rapidly.
How rapidly this occurs must depend critically on the length of the flux tube,
for example. Coronal flux tubes no more than a few Mm in length can be highly populated with
dense thermal plasma within 30~seconds or so, whereas longer flux tubes  would require several minutes to do so.

In the case of the flare of 14 August 2004 the decay of the 17~GHz and 34~GHz 
emission following the initial rise is quite rapid. 
This suggests that relativistic electrons are either injected predominantly into 
the loss cone of the magnetic flux tube at the outset or that trapped electrons 
not initially injected into the loss cone are scattered into it rapidly, which 
could enhance the seismic emission.
The magnetic extrapolation of the region suggests that the field lines connecting 
the photospheres in the neighborhood of the seismic source to their conjugate 
footpoints are indeed short, only a few Mm in length.
This may explain both the rapid and complete decrease in synchrotron emission 
following the impulsive onset and the occurrence of a relatively strong sudden 
white-light signature, and may help to explain a commensurate, relatively strong 
seismic transient emitted from a flare that otherwise is relatively weak.

\end{article}
\end{document}